\documentclass[pdflatex,sn-nature]{sn-jnl}

\usepackage{graphicx}\usepackage{multirow}\usepackage{amsmath,amssymb,amsfonts}\usepackage{amsthm}\usepackage{mathrsfs}\usepackage[title]{appendix}\usepackage{xcolor}\usepackage{etoolbox}
\usepackage{textcomp}\usepackage{manyfoot}\usepackage{booktabs}\usepackage{makecell}\usepackage{algorithm}\usepackage{algorithmicx}\usepackage{algpseudocode}\usepackage{listings}\usepackage{longtable}
\usepackage{tabularx}
\usepackage{booktabs}
\usepackage{siunitx}
\usepackage{subfigure}
\usepackage{bm}

\usepackage{braket}
\usepackage[normalem]{ulem}
\usepackage{xspace}
\usepackage{xcolor}
\usepackage{comment}
\usepackage[capitalise]{cleveref}
\usepackage{tgtermes}

\newcolumntype{C}[1]{>{\centering\arraybackslash}p{#1}}

\newcommand{\s}{\text{s}}
\newcommand{\Hz}{\text{Hz}}
\newcommand{\kHz}{\text{kHz}}
\newcommand{\GeV}{\text{GeV}}

\newcommand{\cm}{\text{cm}}

\newcommand{\km}{\text{km}}

\newcommand{\eq}[1]{Eq.~\eqref{#1}}

\newcommand{\fig}[1]{Fig.~\ref{#1}}
\newcommand{\tab}[1]{Tab.~\ref{#1}}

\def\gw{GW\xspace}

\def\gws{GWs\xspace}

\def\gw{GW\xspace}

\def\gws{GWs\xspace}

\def\uldm{ultralight DM\xspace}

\def\ifos{interferometers\xspace}

\def\ep{\emph{BSD-excess-power}\xspace}

\newcommand{\TFFT}{T_\text{FFT}}

\newcommand{\Tobs}{T_\text{obs}}

\newcommand{\be}{\begin{equation}}
\newcommand{\ee}{\end{equation}}

\newcommand{\Tcoh}{T_{\text{coh}}}
\newcommand{\Lcoh}{L_{\text{coh}}}

\newcommand{\ev}{\,\mathrm{eV}}

\newcommand{\mpl}{M_{\text{Pl}}}

\newcommand{\ccTotalOutliers}{\ensuremath{3233}\xspace}
\newcommand{\ccBandOutliers}{\ensuremath{3197}\xspace}
\newcommand{\ccNonBandOutliers}{\ensuremath{36}\xspace}
\newcommand{\ccSinglePeakVeto}{\ensuremath{21}\xspace}
\newcommand{\ccCutTwoRemain}{\ensuremath{15}\xspace}
\newcommand{\ccLowReSNRVeto}{\ensuremath{7}\xspace}
\newcommand{\ccNonGaussianVeto}{\ensuremath{7}\xspace}
\newcommand{\ccGaussianBelowTh}{\ensuremath{1}\xspace}
\newcommand{\ccCutThreeRemain}{\ensuremath{0}\xspace}

\newcommand{\ccGaussianReSNR}{\ensuremath{-5.15}\xspace}

\newcommand{\bsdNcandtot}{\ensuremath{66876}\xspace} 
\newcommand{\bsdNcoin}{\ensuremath{5519}\xspace} 

\newcommand{\bsdNcoinCRnotdiff}{\ensuremath{5363}\xspace} 

\newcommand{\bsdNcoinoverCRthr}{\ensuremath{0}\xspace}

\newcommand{\bsdCRthroneperc}{\ensuremath{5.53}\xspace}

\theoremstyle{thmstyleone}

\theoremstyle{thmstyletwo}

\theoremstyle{thmstylethree}

\makeatletter
\patchcmd{\@maketitle}
  {\ifequalcont{\par$^{\dagger}$\@equalconttext\par}\fi}
  {}
  {}
  {\PackageError{sn-article-v4}{Could not relocate the equal-contribution note}{}}
\patchcmd{\@maketitle}
  {\ifnum\emailcnt>0\relax}
  {\ifequalcont{\par$^{\dagger}$\@equalconttext\par}\fi
   \ifnum\emailcnt>0\relax}
  {}
  {\PackageError{sn-article-v4}{Could not insert the equal-contribution note}{}}
\makeatother

\begin{document}

\renewcommand*{\theHfigure}{\arabic{section}.\arabic{figure}}
\renewcommand*{\theHtable}{\arabic{section}.\arabic{table}}

\title{Opening the high-frequency band of LIGO for persistent gravitational-wave and dark-matter searches}

\author[1,2]{\fnm{Mai} \sur{Qiao}}
\equalcont{These authors contributed equally to this work.}

\author[3]{\fnm{Alexandre} \sur{G\"ottel}}
\equalcont{These authors contributed equally to this work.}

\author[4]{\fnm{Xuanye} \sur{Fan}}

\author[5,2]{\fnm{Xing-Yu} \sur{Chen}}

\author[6]{\fnm{Pia} \sur{Astone}}

\author[11]{\fnm{Stefano} \sur{Dal Pra}}

\author[7]{\fnm{Evan} \sur{Goetz}}

\author*[1,2]{\fnm{Huai-Ke} \sur{Guo}}
\email{guohuaike@ucas.ac.cn}

\author*[1,2]{\fnm{Andrew L.} \sur{Miller}}
\email{andrew.miller.ligo@ucas.ac.cn}

\author[6,12]{\fnm{Lorenzo} \sur{Silvestri}}

\author[8,9]{\fnm{Fengwei} \sur{Yang}}

\author*[10]{\fnm{Yue} \sur{Zhao}}
\email{zhaoyue@ust.hk}

\affil[1]{\orgname{International Centre for Theoretical Physics Asia-Pacific (ICTP-AP), University of Chinese Academy of Sciences (UCAS)}, \orgaddress{\city{Beijing}, \postcode{100190}, \country{China}}}

\affil[2]{\orgname{Taiji Laboratory for Gravitational Wave Universe, University of Chinese Academy of Sciences}, \orgaddress{\city{Beijing}, \postcode{100049}, \country{China}}}

\affil[3]{\orgdiv{School of Mathematical Sciences}, \orgname{University of Nottingham}, \orgaddress{\city{Nottingham}, \postcode{NG7 2RD}, \country{United Kingdom}}}

\affil[4]{\orgdiv{Department of Physics and Astronomy}, \orgname{Stony Brook University}, \orgaddress{\city{Stony Brook}, \state{NY}, \postcode{11794}, \country{USA}}}

\affil[5]{\orgdiv{Department of Physics}, \orgname{University of Chinese Academy of Sciences (UCAS)}, \orgaddress{\city{Beijing}, \postcode{100190}, \country{China}}}

\affil[6]{\orgname{INFN, Sezione di Roma}, \orgaddress{\city{Roma}, \postcode{I-00185}, \country{Italy}}}

\affil[7]{\orgname{University of British Columbia}, \orgaddress{\city{Vancouver}, \state{BC}, \postcode{V6T 1Z4}, \country{Canada}}}

\affil[8]{\orgdiv{Department of Physics and Astronomy}, \orgname{University of Notre Dame}, \orgaddress{\city{South Bend}, \state{IN}, \postcode{46556}, \country{USA}}}

\affil[9]{\orgname{Johannes Gutenberg-Universit\"{a}t Mainz}, \orgaddress{\postcode{55099}, \city{Mainz}, \country{Germany}}}

\affil[10]{\orgdiv{Department of Physics and Jockey Club Institute for Advanced Study}, \orgname{The Hong Kong University of Science and Technology}, \orgaddress{\city{Hong Kong S.A.R.}, \country{P.R. China}}}

\affil[11]{\orgdiv{INFN-CNAF - Bologna,} \orgaddress{Viale Carlo Berti Pichat, 6/2, 40127 \city{Bologna}, \country{Italy}}}

\affil[12]{\orgname{Universit`a di Roma La Sapienza}, \orgaddress{\city{Roma}, \postcode{I-00185}, \country{Italy}}}

\abstract{
The high-frequency data of gravitational-wave interferometers offer
opportunities to search for weak, persistent signals, but narrow
instrumental features can obscure their interpretation.
Here we characterize the $2$--$5~\mathrm{kHz}$ band of LIGO data from the
first part of the fourth observing run and demonstrate its scientific
reach through a search for ultralight dark matter. We identify higher
harmonics of suspension resonances and features associated with optical
controls, map the affected frequency intervals, and catalogue strong
single-detector features. Three complementary analyses test inter-detector
coherence, finite temporal coherence and spectral consistency for spin-0
dilatons, spin-1 dark photons and spin-2 massive gravitons. No candidate
remains consistent with the assumed dark-matter signals after follow-up.
Our $95\%$ confidence upper limits on dilaton couplings are comparable to
or stronger than existing GEO600 bounds, while our strongest
dark-photon and massive-graviton limits improve on constraints from other experiments by approximately one and three orders of
magnitude, respectively. These results
establish a practical framework for persistent searches in high-frequency
LIGO data and provide an instrumental reference for future searches for
long-lived gravitational waves and fundamental fields.}


\maketitle

\section{Introduction}

Laser-interferometric gravitational-wave detectors have opened a new observational era, revealing phenomena that were previously inaccessible to astronomy~\cite{LIGOScientific:2016aoc,LIGOScientific:2017vwq,LIGOScientific:2021qlt}. Yet the information contained in these instruments has not been exploited uniformly across their observing band. In particular, the high-frequency band above 2000 Hz~\cite{KAGRA:2021tnv,KAGRA:2025yfg,LIGOScientific:2017fdd,LIGOScientific:2020aai,LIGOScientific:2025dup} remains insufficiently characterized for persistent searches, limiting the physics that can be probed. Rapidly rotating neutron stars, boson clouds around black holes and
slowly evolving compact binaries~\cite{Riles:2022wwz,LIGOScientific:2021rnv,KAGRA:2025xvh,Miller:2024jpo,Guo:2022sdd,Bhaumik:2025vlb,Wang:2025lpj,Wang:2025fmh,Arvanitaki:2014wva} can all emit \gws at kilohertz frequencies, but without a fundamental understanding of the noise in this frequency band, searches for such objects cannot be performed.

Throughout the 10-5000 Hz frequency band but especially above 2 kHz, a common obstacle
is the presence of narrow instrumental features, which can resemble
physical signals and accumulate apparent significance over the
observation time. For example, in the $2$--$5~\mathrm{kHz}$ LIGO band, suspension
resonances and structures associated with optical controls would obscure weak, persistent signals and produce candidates with seemingly high significances, reducing our search's sensitivity ~\cite{Capote:2024rmo,LIGO:2024kkz,LSC:2018vzm,O4LIGODetector:2026okh,LIGO:2021ppb}.
Systematic characterization of this environment is therefore needed
to exploit high-frequency data for searches for long-lived
gravitational waves and fundamental fields~\cite{M:2026yut}.

Opening the high-frequency LIGO band to persistent searches requires
an experimental reference that connects the observed power spectrum of the strain data to the
interpretation of potential signals. Here we establish this connection
through a systematic characterization of the dominant instrumental
structures across $2$--$5~\mathrm{kHz}$. We map the disturbed frequencies in each detector  and use this information to rule out potential signals recovered in our searches as astrophysical. The resulting
inventory of cataloged spectral features provides a shared foundation for searches with
different signal models. As an example, we use ultralight dark matter to demonstrate
the scientific potential of this characterization, even though our work here
supports a  broader exploration of long-lived gravitational-wave and
fundamental-field signals.

Wave-like ultralight dark matter (ULDM) provides a physically
well-defined benchmark for this study~\cite{LIGOScientific:2021ffg,LIGOScientific:2025ttj,Gottel:2024cfj,Miller:2025yyx,Delgado:2026hnx,Aiello:2021wlp,Miller:2023kkd,KAGRA:2024ipf}.
In a virialized Galactic halo, the field produces a narrow spectrum
with a finite coherence time set by the velocity distribution.
Across the mass range probed here, its coherence length exceeds the
separation between the LIGO Hanford and Livingston detectors, allowing
correlated responses. In fact, of all the persistent signals mentioned previously, ULDM presents a particularly significant challenge to separate from noise disturbances because their signal morphologies can closely resemble those of stationary instrumental lines. This implies that our noise characterization studies are broadly applicable to searches for which the signal model deviates more substantially from noise disturbances, e.g. for spinning-down, Doppler-shifted \gws from non-axisymmetric, rotating neutron stars.

We consider three representative interactions:
dilatons change the material and optical properties of detector
components~\cite{Gottel:2024cfj,Grote:2019uvn}, dark photons exert forces on the test
masses~\cite{Pierce:2018xmy,Guo:2019ker,Morisaki:2020gui},  and massive
spin-2 fields generate oscillatory metric
perturbations~\cite{Manita:2023mnc,Aoki:2016zgp,Armaleo:2020efr}. Their distinct responses, together
with the expected spectral and temporal behaviour, provide complementary
ways to test whether an observed power excess at a given frequency is compatible with the assumed
dark-matter signal.

Here we characterize the $2$--$5~\mathrm{kHz}$ instrumental landscape
using data from the first part of the fourth observing run (O4a) of the
LIGO Hanford and Livingston detectors. We map the frequency intervals
affected by higher suspension violin-mode harmonics and output-mode-cleaner
structures~\cite{LIGO-T2600523}. We also catalog strong single-detector noise features
and apply three complementary methods to search for ULDM:
cross-correlation~\cite{Guo:2019ker},
Band Sampled Data (BSD)~\cite{Miller:2020vsl} and
Logarithmic Power Spectral Density (LPSD)~\cite{Gottel:2024cfj}.
Each method independently analyzes the data and tests coherence between detectors, and the extent to which potential signals exhibit finite coherence times, and have power spectra that match the expected one from ULDM. However,
no candidate remains that is consistent
with the assumed ULDM signals. The resulting limits demonstrate
competitive sensitivity to scalar dark matter and extend the parameter
space probed for dark photons and massive gravitons. The detector-specific
spectrum characterization provides a reference for future searches, while our application to ULDM demonstrates how potential signals returned by the search can be vetoed through comparison with noise disturbances.

\section{High-frequency detector characterization}
\label{sec:highfreq-characterization}

\begin{figure}[ht!]
    \centering
    \includegraphics[width=\linewidth]{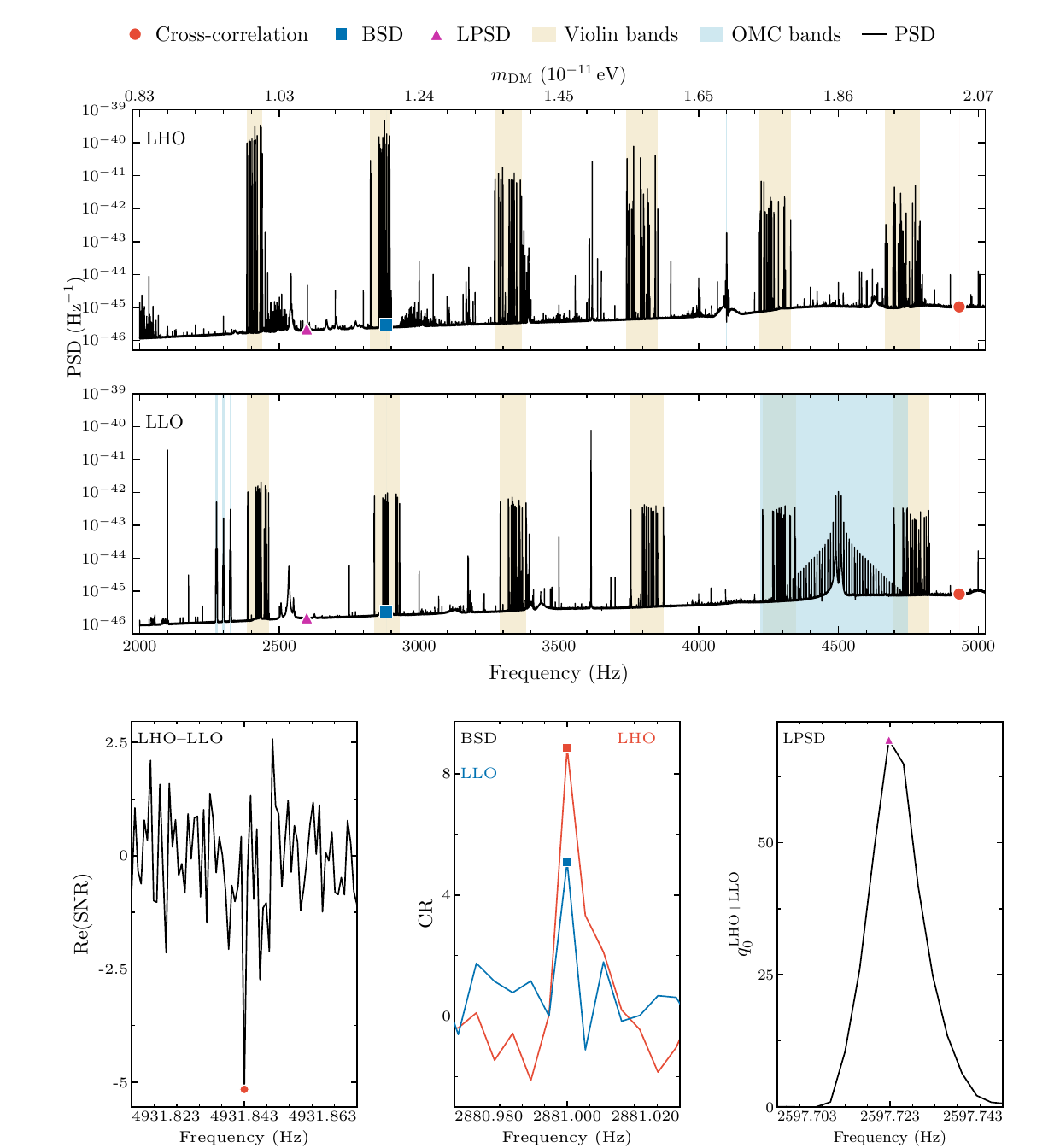}
    \caption{
\textit{Upper:}
O4a PSD spectra (black-solid) for LHO and LLO.
Orange and cyan shaded bands indicate frequency intervals affected by violin-mode and OMC noise.
The red dot, blue squares and purple triangles mark the candidates selected for follow-up in the cross-correlation, BSD and LPSD analyses, respectively.
\textit{Lower:}
Zoomed-in spectra around the three marked candidates.
}
    \label{fig:veto_summary}
\end{figure}

Suspension violin-mode harmonics and structures associated with optical
controls form two prominent families of contamination in the
$2$--$5~\kHz$ LIGO band (\fig{fig:veto_summary}). We map these features
separately in the Hanford (LHO) and Livingston (LLO) detectors using
calibrated strain data from O4a, collected from 24 May 2023 to
16 January 2024. The $5~\kHz$ upper boundary follows the validity range
of the O4a strain calibration~\cite{LIGOScientific:2025snk}.
Data selection and preprocessing are described in
Section~\ref{sec:data-detchar}.

The first family is associated with the harmonics (violin modes) of the suspension systems in LIGO. We map the fifth through tenth harmonics  with fundamental frequencies near $500~\Hz$, which builds upon
the vetted O4a fourth-harmonic bands that provided a
reference for the peak patterns~\cite{LIGO-T2400204}. We locate the band
edges using higher-frequency companion peaks at LHO and pairs of peaks
flanking the central cluster at LLO~\cite{LIGO-T2600523}
(\tab{tab:violin_bands_vertical}).

The second family is associated with the output mode cleaner (OMC),
which filters the output light before photodetection. Dithers used to control the cavity length and beam alignment
produce spectral lines with sidebands~\cite{aLIGO:2020wna,Capote:2024rmo}.
Using a lower-frequency O4a dither line as a
reference~\cite{LIGO-T2400204}, we bound one LHO cavity-length band and
three LLO alignment bands by their outermost identifiable symmetric
sideband pairs. At LLO, the $4500.1~\Hz$ cavity-length dither produces a
comb with approximately $9.6~\Hz$ spacing. Tracing it independently on
each side gives a band of $4220$--$4750~\Hz$
(\fig{fig:asd_omc} and \tab{tab:OMC_bands}).

The resulting maps are consistent with a strong concentration of search outliers
within the identified instrumental bands. Together, 
detector-specific violin-mode and OMC intervals span $1071.5~\Hz$,
or $35.7\%$ of the $2$--$5~\kHz$ band, which is a significant amount of noise pollution that, before this work, hindered searches in this band. We provide the violin-mode and OMC band catalog for each detector and the criteria we used to identify these features  in Section~\ref{sec:volin+omc} and
Ref.~\cite{LIGO-T2600523} to aid in future analyses.  
We demonstrate the utility of our noise characterization efforts through a search for persistent signals from ULDM.

\section{Ultralight dark matter as a benchmark}
\label{sec:dm_models}
Ultralight dark matter provides a benchmark for testing the scientific
reach of the characterized band. Its large occupation number within
a de Broglie volume allows Galactic ULDM to be described as a classical
field, modelled as a superposition of waves with frequencies near
$f_0=m_{\text{DM}}/(2\pi)$, where we use natural units with
$c=\hbar=1$.
The $2$--$5~\kHz$ band therefore
corresponds to masses of approximately
$8.3\times10^{-12}$--$2.1\times10^{-11}~\ev$.
For a virialized Galactic halo, the velocity dispersion produces a
characteristic fractional linewidth of order $10^{-6}$ and a finite
coherence time $\Tcoh=1/\Delta f$. The coherence length is of order
$10^5~\km$ across this band, much larger than the approximately
$3000~\km$ separation between LHO and LLO. The field is therefore
spatially coherent across the two sites and can produce correlated
detector responses~\cite{Derevianko:2016vpm,Pierce:2018xmy,Guo:2019ker,Nakatsuka:2022gaf}.  Such a signal is almost monochromatic, meaning that its morphology can be identical to noise disturbances characterized in Sec. \ref{sec:highfreq-characterization}, further motivating the need for our high-frequency noise characterization.

Ultralight dark matter candidates can be broadly classified by their spin, which can simultaneously be probed with \gw \ifos through their
couplings to Standard Model fields. We consider three representative
bosonic models with distinct spins and interaction mechanisms:
spin-0 dilatons, spin-1 dark photons (DPs) and spin-2 massive gravitons
(MGs). Their interactions are described by
\begin{equation}
\mathcal{L}_{\text{int}}
=
\left\{
\begin{aligned}
&
\frac{\phi}{\Lambda_\gamma}
\frac{F_{\mu\nu}F^{\mu\nu}}{4}
-
\frac{\phi}{\Lambda_{\mathrm{e}}}
m_{\mathrm{e}}
\bar{\psi}_{\mathrm{e}}
\psi_{\mathrm{e}},
&& \text{dilaton},
\\[6pt]
&
-\epsilon e J^\mu A_\mu,
&& \text{DP},
\\[6pt]
&
-\frac{\alpha}{2\mpl}
\chi_{\mu\nu}T^{\mu\nu},
&& \text{MG}.
\end{aligned}
\right.
\label{eq:L_int}
\end{equation}
Here $\phi$, $A_\mu$ and $\chi_{\mu\nu}$ denote the dilaton,
dark-photon and massive-graviton fields, respectively.
The quantities $\psi_e$, $m_e$, $F_{\mu\nu}$ and $T_{\mu\nu}$ denote
the electron field, electron mass, electromagnetic field-strength
tensor and energy--momentum tensor.
The current $J^\mu$ carries baryon number or baryon-minus-lepton
number, and $e$ is the elementary charge.
The parameters $\Lambda_\gamma$ and $\Lambda_e$ are scalar coupling
scales, whereas $\epsilon$ and $\alpha$ are dimensionless vector and
tensor coupling constants.
The dilaton response of the interferometer constrains the
combination of the dimensionless photon and electron couplings
given by~\cite{Gottel:2024cfj}
\[
\lvert d_e+d_{m_e}\rvert
= \sqrt{2}\,\mpl\,
  \bigl\lvert \Lambda_\gamma^{-1}+\Lambda_e^{-1}\bigr\rvert ,
\]
where $\mpl$ is the reduced Planck mass.

These interactions produce distinct interferometer responses.
Dilatons coupled to photons and
electrons~\cite{Damour:1994zq,Stadnik:2014tta,Arvanitaki:2015iga,Damour:2010rp,Fukusumi:2023kqd,Morisaki:2018htj}
modulate the dimensions and refractive indices of the beam splitter
and test masses, producing an optical-path difference between the
arms~\cite{Gottel:2024cfj,Grote:2019uvn,Stadnik:2014tta,Stadnik:2015xbn,Hall:2022zvi}.
Dark photons coupled to baryon number or baryon-minus-lepton number
exert oscillatory forces on the test
masses~\cite{Nelson:2011sf,Arias:2012az,Graham:2015rva,Carney:2019cio,Graham:2015ifn}.
Their strain response includes both differential motion and a
contribution from the finite light-travel time between moving
optics~\cite{LIGOScientific:2021ffg,Morisaki:2020gui,Guo:2019ker,Pierce:2018xmy}.
In ghost-free bigravity, the massive spin-2 field induces an
oscillatory metric perturbation and a differential interferometer
response~\cite{Hassan:2011zd,Manita:2023mnc,Armaleo:2020efr,Marzola:2017lbt}.
The corresponding strain-response expressions, used to translate
the measurements into coupling constraints, are given in
Section~\ref{sec:uldm-response}. 

The shared persistent, narrow-band signature allows these benchmarks
to be investigated with complementary search methods.
Their expected linewidths and finite coherence times guide spectral
and temporal consistency tests, while spatial coherence supports
comparisons between the detectors through their model-dependent
responses. We combine these tests with the instrumental maps to
evaluate candidates and constrain the corresponding couplings.

\section{Ultralight dark matter search strategies}
\label{sec:search-strategies}

We apply the instrumental characterization in
Section~\ref{sec:highfreq-characterization} to a search for ULDM using
three complementary pipelines: cross-correlation (CC), BSD and LPSD.
CC and LPSD cover $2$--$5~\kHz$, while BSD covers
$2$--$4.096~\kHz$.
The methods use different statistics and selection procedures to test
the persistence, coherence and spectral properties of candidate signals.
\tab{tab:candidate_progression} summarizes their three-stage selections
and subsequent follow-up; the detailed implementations are given in
Section~\ref{sec:analysis-pipelines}.

The cross-correlation search measures correlated strain by combining
synchronized Fourier amplitudes from LHO and LLO with noise weighting.
We select \ccTotalOutliers{} initial frequency bins satisfying
$|\mathrm{Re}(\mathrm{SNR})|>5$ or
$|\mathrm{Im}(\mathrm{SNR})|>5$.
The first cut excludes the mapped violin-mode and OMC intervals.
The second rejects spectral peaks confined to one detector.
The third tests for non-Gaussian contamination by cross-correlating
frequency-offset spectra and applies the final significance
requirement~\cite{Guo:2019ker}.
No candidate passes the full selection, which requires
$|\mathrm{Re}(\mathrm{SNR})|>5.9$, corresponding to a false-alarm probability of approximately 1\%.

We additionally examine the sub-threshold feature at
$4931.8433~\Hz$, which passes the instrumental and contamination vetoes
but has $\mathrm{Re}(\mathrm{SNR})=\ccGaussianReSNR$.
Its cumulative-SNR evolution does not by itself reject the ULDM
interpretation (\fig{fig:snr-tobs}). A spectral comparison, however,
finds a narrower profile than expected for the assumed virialized
field (\fig{fig:veto_summary}). This provides an additional
consistency test beyond the discovery threshold. 

The BSD search identifies recurring excess power in each detector,
using Fourier durations matched to the expected ULDM coherence time.
Its number count records how often a frequency bin contains a selected
local power maximum during the observing run; the critical ratio $CR$
measures the excess of this count relative to its background.
We retain \bsdNcandtot{} initial candidates by ranking the counts
within $1~\Hz$ bands.
The three cuts impose the following requirements: frequency coincidence within one bin; critical ratios differing by no more than a factor of five; and  $CR>$ \bsdCRthroneperc, corresponding to a 1\% false-alarm probability, in both detectors. No candidate satisfies the final requirement.

Relaxing the BSD threshold to $CR>5$ retains the feature at
$2881~\Hz$ for further study.
Its critical ratio increases with observing time and continues to
strengthen when $\TFFT$ exceeds the expected field coherence time
(\fig{fig:cr-tfft-tobs}). The latter behaviour is inconsistent with
the finite-coherence prediction for the assumed halo model and is
consistent with a persistent monochromatic instrumental
feature~\cite{Miller:2020vsl,Miller:2022wxu,Vermeulen:2021epa}.

The LPSD search evaluates persistent power excesses through their
spectral shape and relative detector response.
We convert the calibrated strain to equivalent DM-induced strain and
construct frequency-adaptive spectra with Fourier durations matched
to the expected coherence
time~\cite{Gottel:2024cfj,Gottel:2026smd}.
A profile-likelihood statistic $q_0$ measures positive excesses above
a locally fitted background~\cite{Cowan:2010js}.
The threshold $q_0>5.41$, corresponding to a $1\%$ false-alarm
probability, selects 7634 candidate bins.
The first cut groups adjacent bins into 1085 clusters.
The second requires non-zero and mutually compatible detector
responses, leaving 11 candidates.
The third inspects local noise, nearby lines, spectral shape and the
relative detector amplitudes, leaving one feature at
$2597.723~\Hz$.

This LPSD feature only slightly exceeds the local threshold in LHO,
and its peak shape and relative amplitudes pass the preceding checks.
Its growth with accumulated observing time is compatible with the
expected persistence in LLO, while the LHO data are insufficient to
establish that behaviour (\fig{fig:lpsd-q0-tobs}).
We therefore examine its coherence-time dependence by varying
$\TFFT$. The statistic does not show the expected decrease when
the Fourier duration exceeds the field coherence time.
This behaviour is inconsistent with the assumed virialized ULDM
signal and rejects the remaining candidate.

The selected outliers and their follow-up assessments are summarized
in \tab{tab:outliers}, with representative spectra shown in
\fig{fig:veto_summary}.
Across their respective search ranges, no candidate survives the
selection and follow-up as a viable signal under the assumed virialized
ULDM model. The instrumental maps and these complementary consistency
tests provide the basis for the coupling constraints presented below.

\begin{table}[ht]
    \centering
    \caption{Candidate progression through the three pipeline-specific
    selection stages and subsequent follow-up. Entries give the numbers
    remaining after each stage. LPSD Cut~1 groups neighbouring bins into clusters.
    The single LPSD feature surviving Cut~3 is rejected by the additional
    coherence-time test. CC and BSD also examine selected sub-threshold
    outliers, which are not counted as passing their discovery selections.}
    \label{tab:candidate_progression}
    \small
    \setlength{\tabcolsep}{6pt}
    \renewcommand{\arraystretch}{1.15}
    \begin{tabular*}{0.85\linewidth}{@{\extracolsep{\fill}}lccccc@{}}
        \toprule
        Method & Initial & Cut 1 & Cut 2 & Cut 3 & After follow-up \\
        \midrule
        CC & \ccTotalOutliers & \ccNonBandOutliers & \ccCutTwoRemain
           & \ccCutThreeRemain & 0 \\
        BSD & \bsdNcandtot & \bsdNcoin & \bsdNcoinCRnotdiff
            & \bsdNcoinoverCRthr & 0 \\
        LPSD & 7634 & 1085 & 11 & 1 & 0 \\
        \bottomrule
    \end{tabular*}
\end{table}

\begin{table}[tbp]
    \centering
    \caption{Selected outliers and follow-up outcomes. The CC and BSD
    features are below their respective final discovery thresholds;
    the LPSD feature survives the initial three selection stages before
    failing the coherence-time test. Statistics are those of the pipeline
    selecting each feature, with the joint complex SNR given for CC and
    individual-detector statistics for BSD and LPSD.}
    \label{tab:outliers}
    \small
    \setlength{\tabcolsep}{6pt}
    \renewcommand{\arraystretch}{1.15}
    \begin{tabularx}{\linewidth}{@{}ccccc>{\raggedright\arraybackslash}X@{}}
        \toprule
        \multirow{2}{*}{Pipeline}
        & $f_0$
        & $m_{\text{DM}}$
        & \multicolumn{2}{c}{Statistic}
        & \multirow{2}{*}{Follow-up outcome} \\
        & (Hz)
        & ($10^{-11}\,\mathrm{eV}$)
        & LLO
        & LHO
        & \\
        \midrule
        CC & 4931.8433 & 2.040
        & \multicolumn{2}{c}{$\mathrm{SNR}=-5.15+1.92i$}
        & Below discovery threshold; narrower spectral profile than
          expected for the assumed ULDM halo. \\
        BSD & 2881 & 1.192
        & $CR=5.09$ & $CR=8.85$
        & Below discovery threshold; narrower spectral profile than expected for the assumed ULDM halo. \\
        LPSD & 2597.723 & 1.074
        & $q_0=60.8$ & $q_0=6.3$
        & Narrower spectral profile than expected for the assumed ULDM halo. \\
        \bottomrule
    \end{tabularx}
\end{table}

\section{Limits on dark-matter couplings}
\label{sec:coupling-limits}

The null results yield $95\%$ confidence upper limits on the couplings
of the three benchmark fields. We obtain these limits from each
pipeline's strain constraints using the responses in \eq{eq:h0},
with the halo and detector-response assumptions described in
Section~\ref{sec:uldm-response}.
\fig{fig:constraints} compares the individual O4a analyses with existing limits. 

For dilatons, the strongest O4a limits are comparable to or stronger than the GEO600~\cite{Vermeulen:2021epa} constraints at the frequencies searched, whereas the displayed bounds derived from weak equivalence principle (WEP) tests~\cite{Hees:2018fpg,Schlamminger:2007ht} remain more stringent.
This is because the dilaton-induced signal is dominated by variations of the beam splitter rather than by changes in the test masses. As a result, the search cannot benefit from the long arm length in the way a differential arm-length signal would~\cite{Gottel:2024cfj}. We note, however, that the constraints derived from the E\"ot-Wash experiment are based on a WEP test. This result is model-dependent. A conformally coupled dilaton field couples universally to all test objects, thus it is not subject to constraints from such measurements \cite{Damour:2010rp}. Conversely, this specific dilaton model nevertheless predicts a signal for GW experiments. It is therefore important to test this model complementarily with GW observations.

For dark photons, at the most sensitive frequencies, these limits improve on the
displayed E\"ot-Wash constraint by approximately one order of magnitude~\cite{Michimura:2020vxn,Schlamminger:2007ht}. The cross-correlation limits are weaker than the LPSD limits. This is because the high-frequency DPDM signal is dominated by the finite-light-travel-time
contribution. For the Hanford--Livingston pair, the relative detector
geometry suppresses the cross correlation of this
contribution, whereas power-spectrum analyses retain the signal power
in each detector~\cite{Morisaki:2020gui,Manita:2023mnc}.
This geometric suppression leads to the weaker CC limits.

For massive gravitons, CC provides the strongest limits on $\alpha$,
improving on the bound from the inverse-square-law (ISL) test by approximately
three orders of magnitude~\cite{Murata:2014nra,Armaleo:2020efr}. 
In this case, the response is dominated by differential motion and
retains a large Hanford--Livingston
overlap~\cite{Manita:2023mnc}.
The different ordering of the pipeline limits across the three models
illustrates the value of analysing both correlations between sites and
excess power within each detector.

\begin{figure}[ht!]
    \centering
    \includegraphics[width=0.32\linewidth]{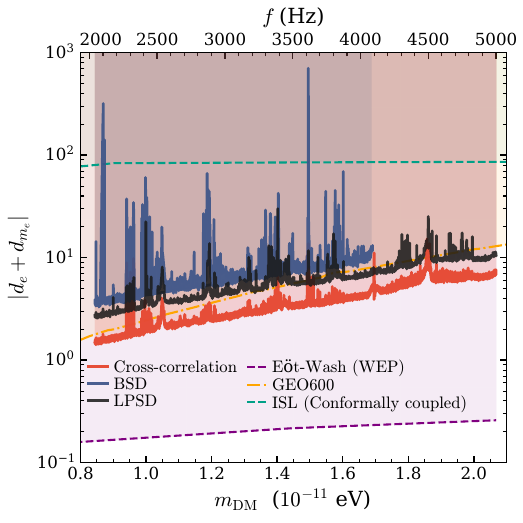}\hfill
    \includegraphics[width=0.32\linewidth]{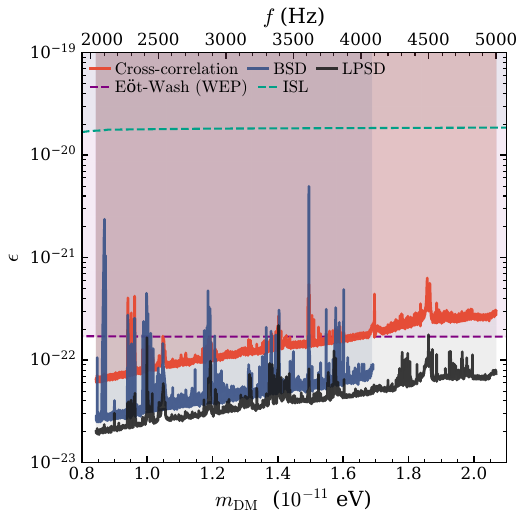}\hfill
    \includegraphics[width=0.32\linewidth]{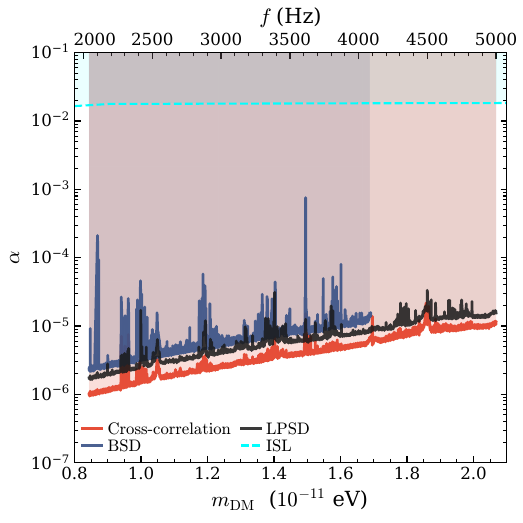}
    \caption{
    \textbf{Constraints on ULDM couplings in the high-frequency band.}
    O4a $95\%$ confidence upper limits on the
    combined dilaton coupling $\lvert d_e+d_{m_e}\rvert$ (left),
    dark-photon coupling $\epsilon$ (middle) and massive-graviton
    coupling $\alpha$ (right).
    Solid red, blue and black curves show the individual CC, BSD and LPSD results, respectively.
    The O4a dark-photon limits assume coupling to baryon number. For all three ULDM candidates,
the Eöt-Wash weak equivalence principle bound  (purple dashed)~\cite{Michimura:2020vxn,Schlamminger:2007ht,Hees:2018fpg}
and the inverse-square-law constraint (cyan dashed)~\cite{Murata:2014nra,Armaleo:2020efr,Fischbach:1999bc} are shown where applicable.
The scalar panel additionally shows the previous GEO600 limit (orange)~\cite{Vermeulen:2021epa}.
}
    \label{fig:constraints}
\end{figure}

\section{Discussion}
\label{sec:discussion}

This work connects a systematic characterization of the
$2$--$5~\kHz$ LIGO O4a noise power spectrum to the selection and interpretation
of persistent-signal candidates.  The detector-specific frequency
maps~\cite{LIGO-T2600523} and the catalogue of strong single-detector
features provide an experimental reference for candidate evaluation at high frequencies.

Our search for ULDM demonstrates the scientific reach of this characterization
through signals with distinct physical couplings and a common
persistent, narrow-band signature. Within their respective search
ranges, the three pipelines find no candidate that survives selection
and follow-up under the assumed virialized-halo model. The resulting
limits provide competitive direct scalar constraints and probe vector
and tensor couplings beyond existing limits.
Persistence alone is insufficient
to identify such a signal because instrumental features can accumulate
significance with observing time as well. As we have shown in the follow-up, spectral width, finite coherence and
the predicted relation between detector responses provide additional
tests of the physical interpretation. 

The noise characterization studies presented here can also inform searches for other
long-lived gravitational-wave signals, including those from rotating
neutron stars, boson clouds and slowly evolving compact
binaries~\cite{Riles:2022wwz,LIGOScientific:2021rnv,Miller:2024jpo}.
Each application requires consistency tests matched to the source's
frequency evolution and its interaction with the detector; however, independently of the signal model, the spectral
characterizations here can guide the treatment of affected frequencies and the follow-up
of candidates, with the signal duration and bandwidth determining how
instrumental features enter the analysis. This makes the
characterization useful well beyond the ULDM searches presented here. 

Further progress can come from longer observations, targeted
instrumental mitigation and continued high-frequency characterization.
Updating the spectral characterizations in subsequent observing runs would track changes
in the detectors and their operating conditions, which can be further contrasted with the signal models expected from ULDM and other persistent signals.
The spectral catalog developed here provides a concrete
starting point for a broader programme of high-frequency searches for
fundamental fields and long-lived gravitational-wave sources.

\appendix
\section{Methods}
\label{sec:methods}
\setcounter{table}{0}
\renewcommand{\thetable}{A\arabic{table}}
\renewcommand{\theHtable}{appendix.\arabic{table}}
\setcounter{figure}{0}
\renewcommand{\thefigure}{A\arabic{figure}}
\renewcommand{\theHfigure}{appendix.\arabic{figure}}

\subsection{Data selection and calibration}
\label{sec:data-detchar}

We analyse calibrated strain data from the LIGO Hanford Observatory
(LHO) and LIGO Livingston Observatory (LLO), collected between
24 May 2023 and 16 January 2024 during O4a. The original time series
are sampled at $16384~\Hz$, corresponding to a Nyquist frequency of
$8192~\Hz$. For these data, the calibration validity determines the
upper boundary of the search. The O4a release specifies a valid
frequency range of $10~\Hz$--$5~\kHz$ within the supplied hourly
calibration-uncertainty envelopes~\cite{LIGOScientific:2025snk}.
We therefore target $2$--$5~\kHz$.

The \texttt{GDS-CALIB\_STRAIN\_CLEAN\_G02} channel includes C00
calibration and self-gating to suppress transient
glitches~\cite{LIGO:2024kkz,LIGO:T2400003}.
We apply Category-1 data-quality vetoes to remove intervals with severe
instrumental problems, leaving $3835.6~\mathrm{h}$ of strain data from
LHO and $3921.3~\mathrm{h}$ from
LLO~\cite{LIGOScientific:2017tza,LIGO:2024kkz,LIGO-T2400058-v3}.
The BSD analysis is restricted to frequencies up to $4096~\Hz$ by
database-construction costs and data-quality considerations.
The data selection specific to each pipeline is described in
Section~\ref{sec:analysis-pipelines}.

\subsection{High-frequency detector characterization}
\label{sec:volin+omc}

We characterize two prominent families of instrumental structure in
the $2$--$5~\kHz$ band: high-order harmonics of the suspension violin
modes and features associated with output-mode-cleaner (OMC) controls.
Their frequencies and spectral extents differ between LHO and LLO,
so we construct separate maps for the two detectors.

\subsubsection{Reference spectra}

We construct power spectral densities (PSDs) separately for LHO and
LLO using short Fourier transforms (SFTs) of duration
$T_{\mathrm{SFT}}=1800~\mathrm{s}$.
The nominal frequency-bin spacing is therefore
$\delta f=1/T_{\mathrm{SFT}}\simeq0.556~\mathrm{mHz}$.
The band maps summarize the structures identified in these aggregate
O4a reference spectra. Each reported band is a bounding interval for
an identified group of features; its width is distinct from the
linewidths of the individual peaks within it.

\subsubsection{Suspension violin-mode harmonics}

Violin modes arise from transverse resonances of the suspension
fibres~\cite{LIGO:2024kkz}. We focus on the violin-mode families with
fundamental frequencies near $500~\Hz$~\cite{Aston:2012ona}. Their harmonics form groups of adjacent peaks with approximately
Lorentzian profiles~\cite{Littenberg:2014oda}. At lower frequencies,
these groups occur near $500n~\Hz$ and shift progressively below this
reference as the harmonic order $n$ increases~\cite{LIGO-T2400204,O4LIGODetector:2026okh,LIGO-G1601163}.

Measurements from around O3 provide frequency references for the
higher harmonics. The LHO measurements predate O3 and extend through
the eighth harmonic~\cite{Sorazu:LHOaLOG46597}. The LLO measurements
are primarily from O3 and extend through the ninth
harmonic~\cite{Sorazu:LLOaLOG45485}.

We identify the fifth through tenth harmonics in the O4a reference
spectra. For each harmonic covered by the historical measurements, we
search from $30~\Hz$ below the lowest reported peak to $30~\Hz$ above
the highest. For the ninth and tenth harmonics at LHO and the tenth at
LLO, we search from $500(n-1)~\Hz$ to $500n~\Hz$. Within these windows,
we look for recurring peak patterns across harmonic orders and
compare them with the fourth-harmonic bands in each detector's O4a
vetted line list~\cite{LIGO-T2400204}. We require each peak used for
band identification to span at least ten frequency bins, equivalent to
approximately $5.6~\mathrm{mHz}$ for an SFT duration of $1800~\mathrm{s}$.

At LHO, we call the central peak of each resolved feature the main
peak and the other resolved peaks in the same feature its companions.
Weaker companions often lie on the high-frequency side of the main
peak, a pattern that recurs in all six high-frequency bands. The
lowest- and highest-frequency groups exhibiting this pattern determine
the band boundaries. The line catalogue~\cite{LIGO-T2600523} reports the
companion-peak statistics.

At LLO, the fourth-harmonic band provides a reference morphology
with a dense central cluster and a distinct pair of peaks on each
side. Both edge pairs are resolved in the fifth, sixth, eighth, ninth
and tenth harmonics. Each pair is typically more than $6~\Hz$ from the
nearest interior peak group. The current catalogue gives separations
of $1.12$--$1.43~\Hz$ within the low-frequency pairs and
$0.66$--$2.69~\Hz$ within the high-frequency
pairs~\cite{LIGO-T2600523}. The outer peak of each pair defines the
corresponding band boundary. Overlap with another spectral feature
makes the upper boundary of the seventh harmonic less certain.
We report the resulting detector-specific intervals at integer-Hz
precision in \tab{tab:violin_bands_vertical}.

\begin{table}[htbp]
\caption{Frequency intervals of the fifth-to-tenth suspension
violin-mode harmonics identified in the O4a PSDs.}
\label{tab:violin_bands_vertical}
\centering
\small
\setlength{\tabcolsep}{6pt}
\renewcommand{\arraystretch}{1.15}
\begin{tabular*}{0.85\linewidth}{@{\extracolsep{\fill}}ccc@{}}
\toprule
Harmonic & LHO (Hz) & LLO (Hz) \\
\midrule
$n{=}5$  & 2383--2440 & 2385--2463 \\
$n{=}6$  & 2825--2897 & 2838--2932 \\
$n{=}7$  & 3269--3368 & 3289--3383 \\
$n{=}8$  & 3741--3854 & 3756--3875 \\
$n{=}9$  & 4217--4330 & 4228--4347 \\
$n{=}10$ & 4666--4792 & 4698--4826 \\
\bottomrule
\end{tabular*}
\end{table}

\subsubsection{Output-mode-cleaner structures}

The OMC is an optical filter cavity between the interferometer's
antisymmetric port and the output photodiodes. It transmits the
optical mode carrying the differential-arm signal while rejecting
higher-order transverse modes and modulation
sidebands~\cite{aLIGO:2020wna,Fricke:2011dv,Capote:2024rmo}.
The OMC uses small periodic modulations, or dithers, to sense
deviations from its operating point. A piezoelectric actuator on an OMC
mirror modulates the cavity length, while angular dithers of the
steering mirrors OM1 and OM3 modulate the alignment of the incident
beam. In the strain spectrum, a dither can produce a narrow line at
the applied frequency, accompanied by broad or narrow sidebands that
are approximately symmetric about that line.

In the lower-frequency O4a spectrum, the vetted LLO line list
identifies a broad symmetric structure extending approximately
$3.84~\Hz$ on either side of a dither line at
$1750.1~\Hz$~\cite{LIGO-T2400204}. This line is associated with the
OM1 pitch dither~\cite{LIGOaLOG:63266}. We use this feature as a
reference when identifying sidebands at higher frequencies.

For the high-frequency O4a analysis, we use the LHO and LLO
commissioning logs to identify the dither frequencies and examine the
strain PSDs around those frequencies. The affected bands and their
instrumental associations are listed in \tab{tab:OMC_bands}, and the
corresponding spectra are shown in \fig{fig:asd_omc}.

The LHO cavity-length dither at $4100~\Hz$ and the three LLO
alignment dithers at $2275.1$, $2300.1$ and $2325.1~\Hz$ each have
broad, approximately symmetric sidebands. We trace each structure
outward from its dither line to the outermost identifiable pair of
matching sidebands, where the structure merges into the local noise
floor. We report the band edges in $0.5~\Hz$ increments.

The LLO cavity-length dither at $4500.1~\Hz$ produces a comb.
We trace the comb independently on each side of the dither line
to the outermost tooth distinguishable from the local noise floor.
Both this assessment and the identification of the symmetric bands
use visual contrast with the adjacent spectrum. The lower noise floor on the low-frequency side
allows us to trace the comb farther below the dither line than above
it. The outermost identifiable teeth are at $4221.89$ and
$4749.52~\Hz$. Rounding the two edges outward to multiples of
$10~\Hz$ gives the interval $4220$--$4750~\Hz$. A linear fit to the
frequencies of 55 identified teeth as a function of their integer
orders gives a spacing of $9.593~\Hz$.

\begin{table}[htbp]
\caption{Frequency intervals associated with OMC dither structures
in the $2$--$5~\kHz$ O4a PSDs.}
\label{tab:OMC_bands}
\centering
\small
\setlength{\tabcolsep}{6pt}
\renewcommand{\arraystretch}{1.15}
\begin{tabular*}{0.85\linewidth}{@{\extracolsep{\fill}}cclcc@{}}
\toprule
Detector & Carrier (Hz) & Instrumental association & O4a band (Hz)
         & aLOG \\
\midrule
LHO & 4100.0 & OMC mirror (cavity length) & 4098.5--4101.5
    & \cite{LIGOaLOG:67588} \\
LLO & 2275.1 & OM1 (yaw)                  & 2271.0--2279.5
    & \cite{LIGOaLOG:33257} \\
LLO & 2300.1 & OM3 (yaw)                  & 2296.0--2304.0
    & \cite{LIGOaLOG:33257} \\
LLO & 2325.1 & OM3 (pitch)                & 2321.0--2329.0
    & \cite{LIGOaLOG:63266} \\
LLO & 4500.1 & OMC mirror (cavity length) & 4220--4750
    & \cite{LIGOaLOG:44177} \\
\bottomrule
\end{tabular*}
\end{table}

\begin{figure}[ht!]
    \centering
    \includegraphics[width=0.32\linewidth]{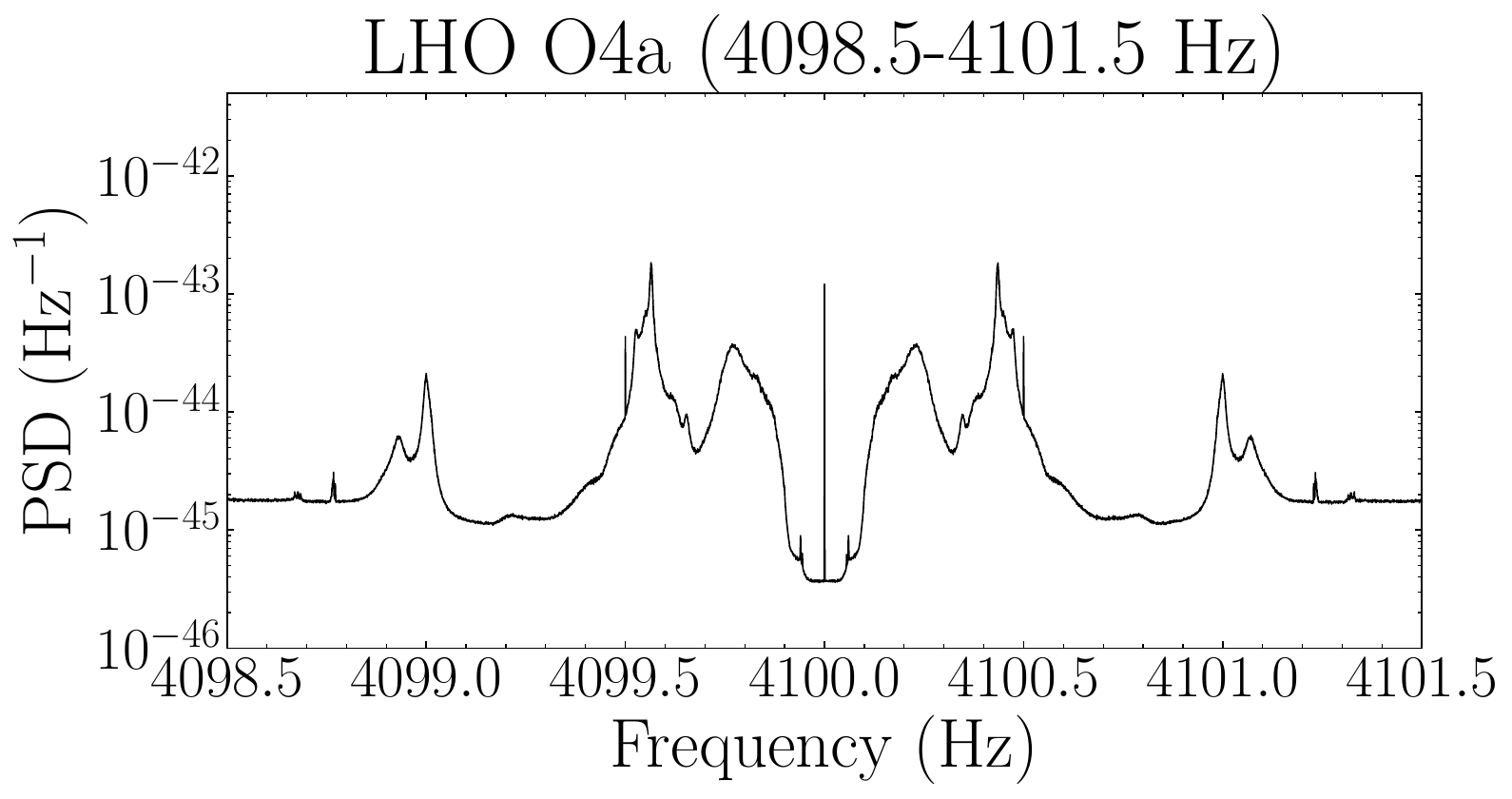}\hfill
    \includegraphics[width=0.32\linewidth]{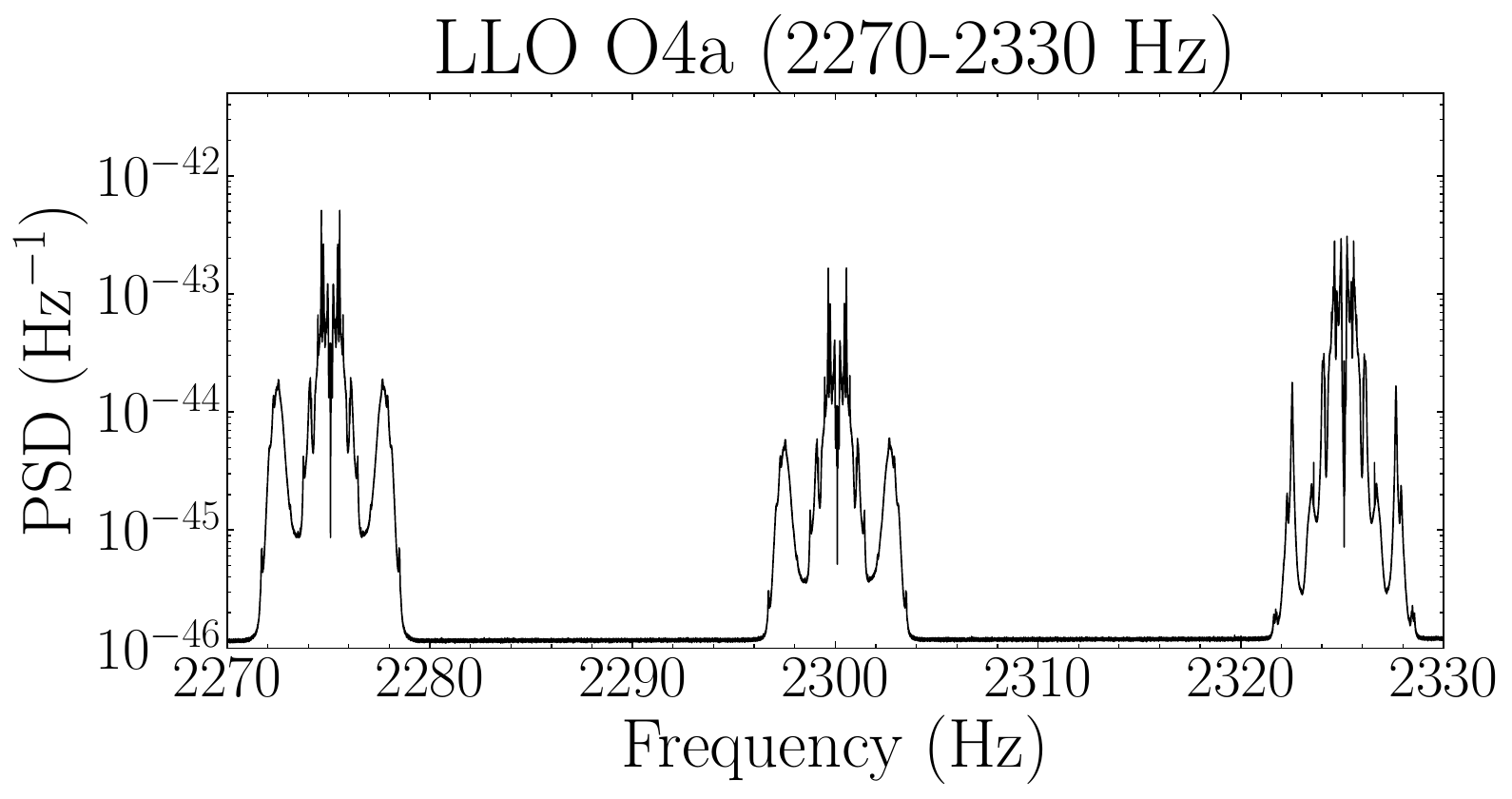}\hfill
    \includegraphics[width=0.32\linewidth]{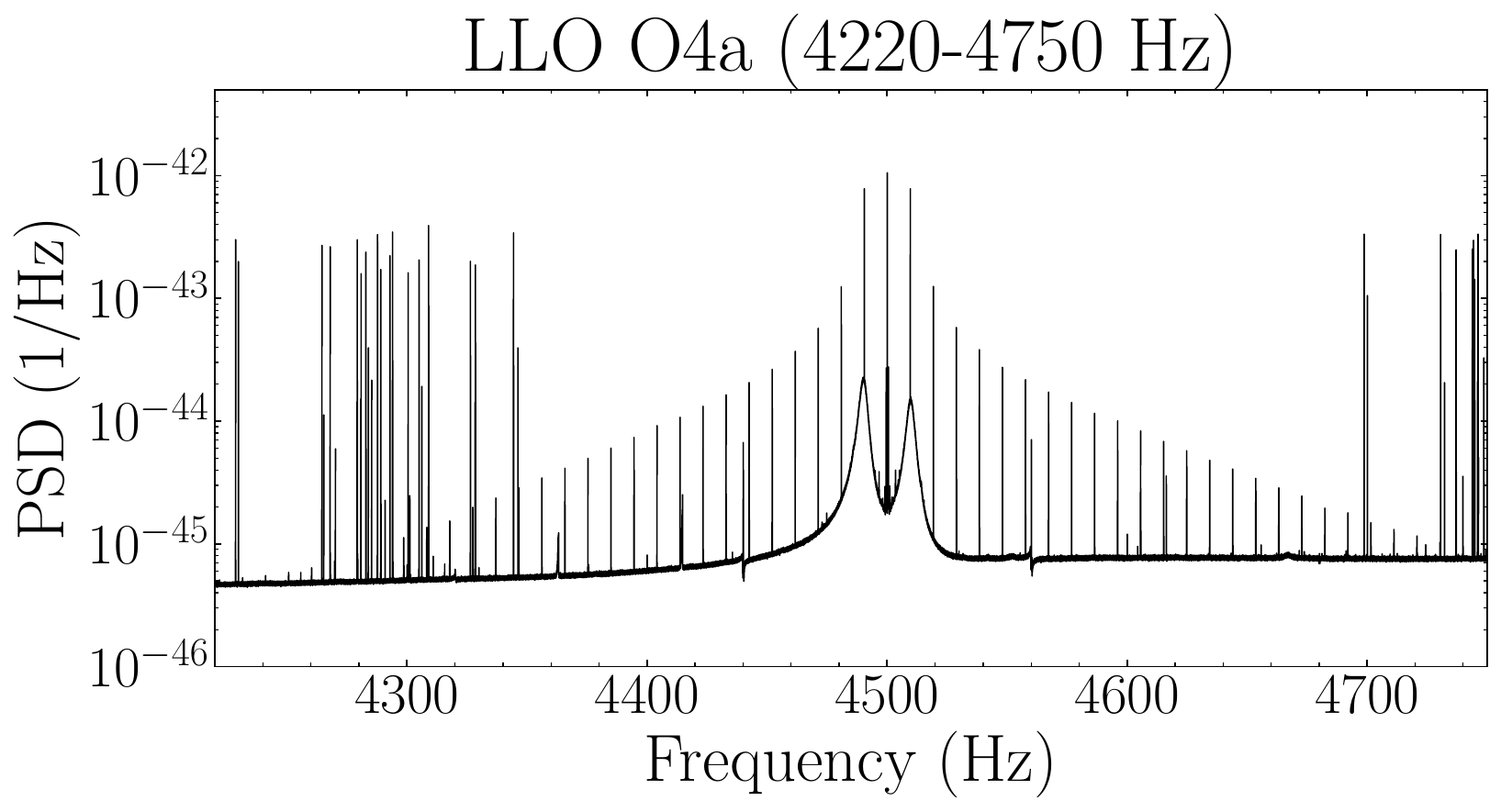}

    \caption{\textbf{OMC-related high-frequency structures.} PSDs show
    the LHO feature near $4100~\Hz$ (left), the LLO angular-dither
    sidebands near $2300~\Hz$ (middle), and the broad LLO comb around
    $4500~\Hz$ (right). The identified frequency intervals are given
    in \tab{tab:OMC_bands}.}
\label{fig:asd_omc}
\end{figure}

\subsection{ULDM signal model and detector responses}
\label{sec:uldm-response}

We describe the benchmark fields in \eq{eq:L_int} as classical
Galactic fields oscillating near $f_0=m_{\text{DM}}/(2\pi)$, using
natural units with $c=\hbar=1$~\cite{Miller:2025yyx}.
The characteristic coherence scales quoted in this work are
\[
\frac{\Delta f}{f_0}\simeq\frac{v_0^2}{2},
\qquad
\Tcoh=\frac{1}{\Delta f},
\qquad
\Lcoh=\frac{1}{f_0v_0},
\]
where $v_0\simeq220~\km/\s$ is the characteristic halo speed.
The corresponding length,
$\Lcoh\simeq1.6\times10^5
(10^{-11}~\ev/m_{\text{DM}})~\km$,
exceeds the LHO--LLO separation throughout the search band.
The field is therefore treated as spatially coherent across the two
sites, with the relation between the measured strains determined by
the detector responses.

{

We define the strain amplitude $h_0$ through the mean-square fractional
arm-length variation. This time-domain correlator is related to the
one-sided strain power spectral density $S_h(f)$ by
\begin{equation}
\left\langle \frac{\Delta L^2(t)}{L^2} \right\rangle
= \int_0^\infty \mathrm{d}f\, S_h(f)
\equiv \frac{1}{2} h_0^2,
\label{eq:dldl}
\end{equation}
where $\Delta L$ and $L$ are the differential and nominal arm length of the detector,
respectively.
}
The strain-amplitude expressions for the three interactions are
\cite{Gottel:2024cfj,LIGOScientific:2021ffg,Morisaki:2020gui,Aoki:2016zgp,Manita:2023mnc}
\begin{equation}
h_0 =
\left\{
\begin{aligned}
&\frac{1}{2\pi}
\left(
    \frac{1}{\Lambda_\gamma}
    +
    \frac{1}{\Lambda_e}
\right)
\frac{
    \sqrt{2\rho_{\text{DM}}}
}{
    f_0 A_{\text{cal}}(f_0)
},
&& \text{dilaton},
\\[6pt]
&\epsilon
C_{\text{geo}}
\frac{q}{M}
\frac{v_{\text{rms}}}{2\pi}
\sqrt{2\rho_{\text{DM}}}
\frac{e}{f_0}
C_{\text{FTE}}(f_0),
&& \text{DP},
\\[6pt]
&\frac{\alpha}{\pi f_0 M_{\text{Pl}}}
\sqrt{
    \frac{\rho_{\text{DM}}}{5}
},
&& \text{MG}.
\end{aligned}
\right.
\label{eq:h0}
\end{equation}
Here 
$\rho_{\text{DM}}=0.4~\GeV/\cm^{-3}$
is the local dark-matter density used in
the benchmark responses~\cite{deSalas:2020hbh}.

For dilatons, $A_{\text{cal}}(f_0)$ is the dimensionless factor
converting gravitational-wave-calibrated strain into the equivalent
fractional expansion of the optical components. It includes the
beam-splitter and test-mass transfer functions~\cite{Gottel:2024cfj}.

For dark photons, $q$ is the charge number carried by a test mass of
mass $M$ for the chosen current, and $e$ is the elementary charge.
The factor $C_{\text{geo}}$ describes the field--detector geometry,
and $v_{\text{rms}}$ is the root-mean-square halo speed.
Combining both the differential arm-length contribution and the finite-light-travel-time contribution, one has
\[
C_{\text{FTE}}(f_0)=
    \sqrt{
    1 + \frac{3}{4} \left( \frac{2\pi f_0 L}{v_{\text{rms}}} \alpha_{IJ} \right)^2},
\]
where $L$ is the arm length. The factor $\alpha_{IJ}$ is introduced
to account for the common-mode overlap of detectors $I$ and $J$,
with unity used for a single detector
\cite{LIGOScientific:2021ffg,Morisaki:2020gui}.
The vector limits reported here assume coupling to baryon number.
We use $q/M=1/m_p$, where $m_p=0.938~\GeV$ is the proton
mass~\cite{ParticleDataGroup:2022pth}, and
$C_{\text{geo}}=\sqrt{2}/3$ for the L-shaped LIGO
detectors~\cite{Pierce:2018xmy}.
We set $\alpha_{IJ}=0.18$ for the LHO--LLO baseline in the
cross-correlation analysis~\cite{LIGOScientific:2021ffg,Morisaki:2020gui,Manita:2023mnc}
and $\alpha_{IJ}=1$ for the single-detector BSD and LPSD analyses.

The massive-graviton response in \eq{eq:h0} is averaged over all
five polarizations~\cite{Manita:2023mnc}.
We use the reduced Planck mass $\mpl=2.435\times10^{18}~\GeV$~\cite{ParticleDataGroup:2022pth}.

\subsection{Search statistics and candidate selection}
\label{sec:analysis-pipelines}

The three pipelines use the data selection and instrumental characterization
described above, with different observables and background models.
CC and LPSD cover $2$--$5~\kHz$, and BSD covers
$2$--$4.096~\kHz$.
Here we define each statistic and its candidate-selection procedure;
\tab{tab:candidate_progression} summarizes the numbers retained at each
stage.
The subsequent candidate follow-up is described in Section~\ref{sec:candidate-followup}.

\subsubsection{Cross-correlation}
\label{sec:pipeline-xcorr}

We select observing times common to LHO and LLO and divide them into
5696 synchronized pairs of $1800~\mathrm{s}$ segments.
After applying a Tukey window, we calculate the short Fourier transforms
(SFTs) and construct a common-frequency cross spectrum.
For each detector and segment, neighbouring non-signal bins provide a
local estimate of the one-sided noise power spectral density
(PSD)~\cite{Guo:2019ker}. These estimates determine the cross-spectrum
weights and statistical uncertainty under the baseline assumption of
independent Gaussian detector noise.
The SFT duration $T_{\mathrm{SFT}}=1800~\mathrm{s}$ gives a frequency
spacing of approximately $0.556~\mathrm{mHz}$.

At frequency bin $j$, we multiply one detector's Fourier coefficient by
the complex conjugate of the other's and combine the synchronized
products with inverse noise-power weighting~\cite{Allen:1997ad}.
For coefficients $\tilde z_{a,ij}$ and powers
$P_{a,ij}=T_{\mathrm{SFT}}\mathrm{PSD}_{a,ij}/2$, the combined spectrum,
component variance and normalized SNR are
\begin{align}
S_j &= \frac{1}{N_{\mathrm{SFT}}}\sum_{i=1}^{N_{\mathrm{SFT}}}
\frac{\tilde z_{1,ij}\tilde z_{2,ij}^{*}}{P_{1,ij}P_{2,ij}}, \label{eq:sj}\\
\sigma_j^2 &= \frac{1}{N_{\mathrm{SFT}}^2}\sum_{i=1}^{N_{\mathrm{SFT}}}
\frac{1}{2P_{1,ij}P_{2,ij}}, \label{eq:sigma2j}\\
\mathrm{SNR}_j &= S_j/\sigma_j. \label{eq:snr_data}
\end{align}
Here $a$, $i$ and $j$ label detector, segment and frequency bin, and
$N_{\mathrm{SFT}}$ is the number of synchronized segment pairs.
The real and imaginary components have the same variance $\sigma_j^2$
in the assumed noise model.
Both components enter the initial candidate and noise diagnostics;
the ULDM response is tested through the real component, interpreted
using the signed response of the relevant model~\cite{Manita:2023mnc}.

We initially select bins satisfying
$|\mathrm{Re}(\mathrm{SNR})|>5$ or
$|\mathrm{Im}(\mathrm{SNR})|>5$, obtaining
\ccTotalOutliers{} candidates.

\emph{Cut~1: instrumental-band masking.}
We exclude candidates within the union of the detector-specific
violin-mode and OMC intervals in
Tables~\ref{tab:violin_bands_vertical} and~\ref{tab:OMC_bands}.
This rejects \ccBandOutliers{} candidates and leaves
\ccNonBandOutliers{} outside the masked intervals.

\emph{Cut~2: single-detector spectral inspection.}
For a candidate at frequency $f_0$, we estimate a local PSD floor
separately in each detector by taking the arithmetic mean within
$f_0\pm0.1~\Hz$ and divide the candidate-bin PSD by this floor.
A peak-to-background ratio above 4 in one detector and below 2 in the
other flags a possible single-detector artifact.
Before applying the veto, we search the latter detector for another
peak within the expected ULDM signal width whose ratio exceeds 4,
using the same local floor. This allows for signal power distributed
over neighbouring bins.
If no such counterpart is present, we veto the candidate.
The cut removes \ccSinglePeakVeto{} candidates and leaves
\ccCutTwoRemain{}.

\emph{Cut~3: significance and frequency-lag controls.}
We adopt $|\mathrm{Re}(\mathrm{SNR})|>5.9$ as the final discovery
threshold. To diagnose non-Gaussian contamination, we also repeat the
cross-correlation with non-zero relative frequency offsets between the
detector spectra. A threshold of 4 on the real or imaginary lagged
SNR within $\pm0.1~\Hz$ flags contaminated control
bins~\cite{Guo:2019ker}.
Of the \ccCutTwoRemain{} candidates entering this stage,
\ccLowReSNRVeto{} have $|\mathrm{Re}(\mathrm{SNR})|<5$, and
\ccNonGaussianVeto{} fail the frequency-lag check.
The remaining \ccGaussianBelowTh{} feature, at $4931.8433~\Hz$,
has $\mathrm{Re}(\mathrm{SNR})=\ccGaussianReSNR$ and is below
the final threshold in magnitude.
Thus \ccCutThreeRemain{} candidates pass the full discovery selection.

\begin{figure}[ht!]
\centering
\includegraphics[width=\linewidth]{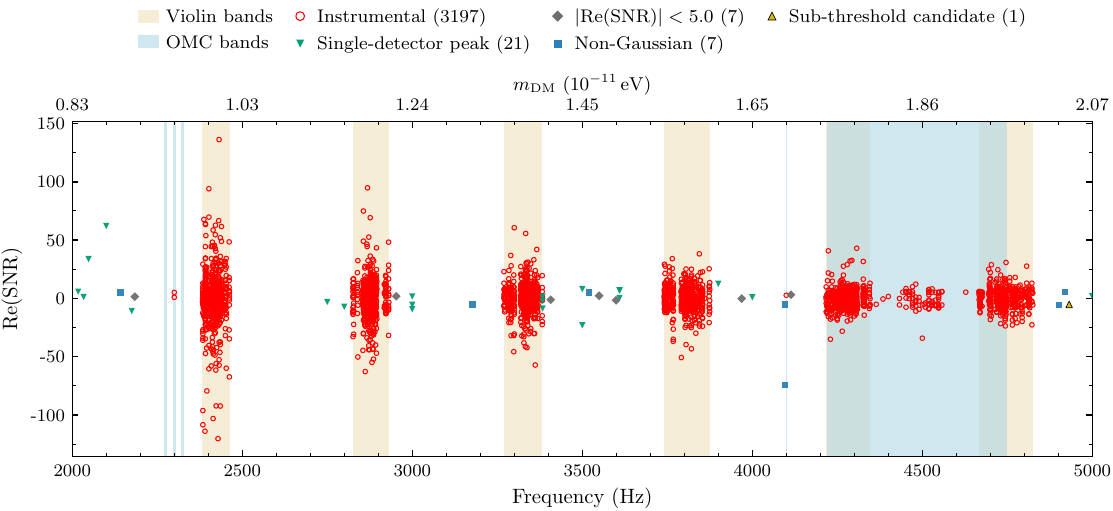}
\caption{\textbf{Cross-correlation candidate vetoes in O4a.}
Orange and cyan bands mark violin-mode and OMC contamination.
Red open circles mark \ccBandOutliers{} candidates in these bands;
green inverted triangles mark \ccSinglePeakVeto{} single-detector
peaks; grey diamonds mark \ccLowReSNRVeto{} candidates with
$|\mathrm{Re}(\mathrm{SNR})|<5$; blue squares mark
\ccNonGaussianVeto{} frequency-lag failures.
The yellow triangle denotes the remaining sub-threshold feature.
The upper axis gives the dark-matter mass in units of
$10^{-11}~\mathrm{eV}$.}
\label{fig:xcorr-veto-summary}
\end{figure}

Figure~\ref{fig:xcorr-veto-summary} summarizes this selection.
The additional follow-up of the sub-threshold feature at
$4931.8433~\Hz$ is described in Section~\ref{sec:candidate-followup}.

\subsubsection{BSD excess power}
\label{sec:pipeline-bsd}

For the Band-Sampled Data (BSD) pipeline, dedicated short-FFT databases
for the high-frequency band are converted to one-month, $10~\Hz$
complex time series~\cite{Astone:2014esa,Piccinni:2018akm,Astone:2005fj}.
We process LHO and LLO separately.
Within each $1~\Hz$ band, we divide the data into chunks of duration
$T_{\mathrm{FFT}}$, chosen to match the expected ULDM coherence
time~\cite{Miller:2020vsl}.
Each chunk is Fourier transformed, and an autoregressive PSD estimator
tracks the slowly varying noise floor without closely following large,
narrow peaks. Normalizing the spectral power by this noise estimate
gives the ratio $R$.
Local frequency maxima with $R>2.5$ form a binary time--frequency
peakmap. Each selected peak contributes unit weight, and its phase is
discarded.

Projecting the peakmap over time gives a number count $n$ for each
frequency bin, measuring the recurrence of selected excess power.
For each detector, the critical ratio is
\begin{equation}\label{eq:bsd-cr}
CR=(n-\mu)/\sigma,
\end{equation}
where $\mu$ and $\sigma$ are the mean and standard deviation of the
projected counts.
The baseline Gaussian-noise model gives a standardized statistic;
instrumental features can generate non-Gaussian tails in real data.

We rank bins by number count within each band of width $B=1~\Hz$.
The number retained in each detector is set using the approximate
accidental-coincidence relation~\cite{Astone:2014esa}
\begin{align}
K_{\mathrm{coin}} &\simeq K_1K_2/K_{\mathrm{tot}},\\
K_{\mathrm{tot}} &= B/\delta f=T_{\mathrm{FFT}}B.
\end{align}
Here $K_1$ and $K_2$ are the retained numbers in LHO and LLO, and
$K_{\mathrm{tot}}$ is the number of frequency bins in the ranking band.
Setting $K_1=K_2=K$ and targeting $K_{\mathrm{coin}}=1$ gives
$K\simeq\sqrt{T_{\mathrm{FFT}}B}$, which varies with the Fourier
duration. This procedure yields \bsdNcandtot{} initial candidates.

\emph{Cut~1: detector coincidence.}
We match the selected LHO and LLO frequencies within one bin,
$\delta f=1/T_{\mathrm{FFT}}$, retaining \bsdNcoin{} coincidences.
Strong non-coincident features also contribute to the instrumental
catalogue: 199 are identified in LHO and 88 in LLO, with their critical
ratios, expected ULDM linewidths and instrumental-band associations.

\emph{Cut~2: statistic consistency.}
We require the single-detector critical ratios to differ by no more
than a factor of five, leaving \bsdNcoinCRnotdiff{} coincidences.

\emph{Cut~3: individual-detector significance.}
The final selection requires $CR>\bsdCRthroneperc$ in both detectors;
\bsdNcoinoverCRthr{} candidates satisfy this requirement.
Relaxing the threshold to $CR>5$ retains the feature at $2881~\Hz$
for additional investigation (\tab{tab:outliers}).

\begin{figure}[ht!]
\centering
\includegraphics[width=\linewidth]{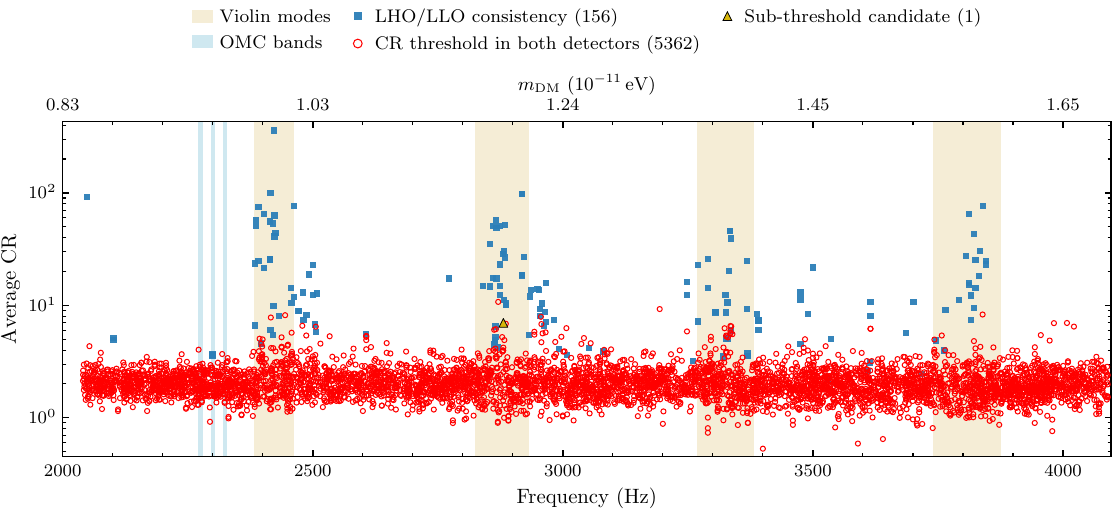}
\caption{\textbf{Candidate selection and veto progression for \ep.}
Ranked candidates are selected in $1~\Hz$ bands and tested for
frequency coincidence, consistency of the single-detector critical
ratios, and the final requirement
$CR>\bsdCRthroneperc$ in both detectors.
The lower threshold $CR>5$ identifies the feature at $2881~\Hz$
for additional follow-up.}
\label{fig:O4a_bsd_progress}
\end{figure}

Figure~\ref{fig:O4a_bsd_progress} summarizes the candidate selection.
The follow-up of the $2881~\Hz$ feature is described in
Section~\ref{sec:candidate-followup}.

\subsubsection{LPSD}
\label{sec:pipeline-lpsd}

For the Logarithmic Power Spectral Density (LPSD) pipeline, we convert
calibrated strain to equivalent DM-induced strain using simulated
gravitational-wave and dark-matter transfer
functions~\cite{Gottel:2024cfj}.
We then construct a frequency-adaptive spectrum, choosing the Fourier
integration time in each bin to optimize the response to the expected
ULDM coherence time~\cite{Gottel:2026smd,trobs_improved_2006}.
A cubic spline in log--log space describes the smooth local PSD
background, and a skew-normal distribution models the residual
fluctuations. Likelihoods for the individual data segments are combined
at each trial frequency.

The signal at each trial frequency is represented by a peak of height
$\eta$, with $\eta=0$ defining the background-only hypothesis.
We fit the signal and background nuisance parameters jointly and
compare the best fit with the background-only fit using the one-sided
profile-likelihood statistic~\cite{Cowan:2010js}
\begin{equation}\label{eq:q0}
q_0=\begin{cases}
-2\ln\dfrac{\mathcal L(0,\hat{\hat{\bm\theta}})}
{\mathcal L(\hat\eta,\hat{\bm\theta})}, & \hat\eta\geq0,\\
0, & \hat\eta<0.
\end{cases}
\end{equation}
Here $\bm\theta$ denotes nuisance parameters; double hats indicate
the background-only fit and single hats the unrestricted best fit.
A negative fitted height is assigned $q_0=0$, so a downward fluctuation
does not count as a positive excess.
The noise distribution is checked at randomly selected frequency bins
against the expected asymptotic behaviour.

We initially select bins with $q_0>5.41$, corresponding to a 1\%
\emph{local} false-alarm probability under the adopted asymptotic
calibration. This gives 7634 candidate bins.

\emph{Cut~1: frequency clustering.}
Adjacent bins above threshold are grouped into a single cluster,
reducing the initial list to 1085 spectral features.

\emph{Cut~2: non-zero and compatible detector responses.}
We require the inferred amplitude in each detector to differ from zero
and test the compatibility of the two inferred amplitudes using a
Student's $t$-test. Both checks use a 1\% significance level.
Their combined application leaves 11 candidates.

\emph{Cut~3: noise and morphology inspection.}
We inspect each candidate neighbourhood for transient excesses not
captured by the spline and for contamination from nearby lines.
We also compare the peak shape and relative LHO--LLO amplitudes with
the ULDM transfer-function predictions.
These checks reject 10 of the 11 candidates, leaving the feature at
$2597.723~\Hz$ (\tab{tab:outliers}).
Its LHO statistic is only slightly above the local threshold.
No obvious nearby excess noise is identified, and neither its peak
shape nor the relative detector amplitudes provide grounds for
rejection at this stage.

\begin{figure}[ht!]
\centering
\includegraphics[width=\linewidth]{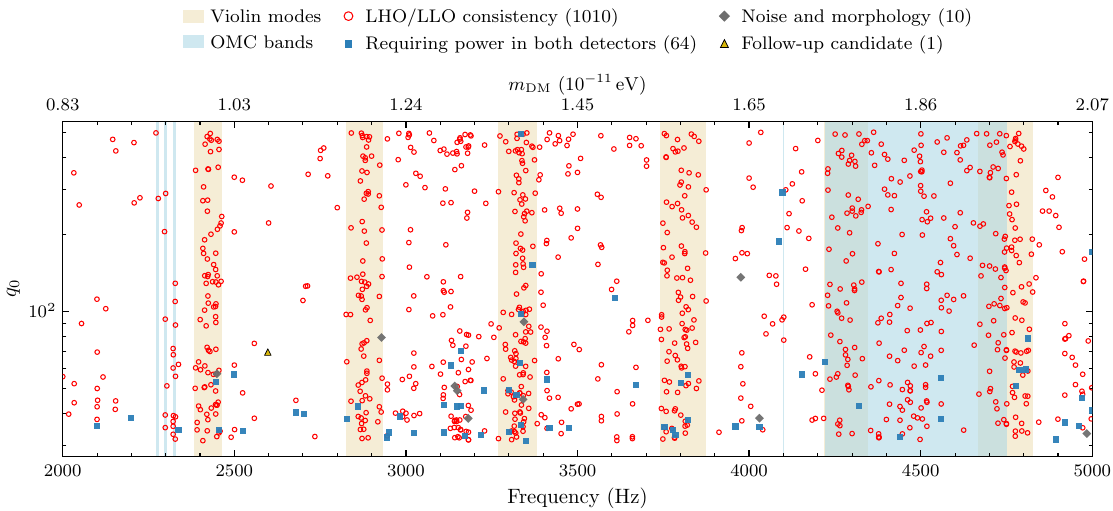}
\caption{\textbf{LPSD candidate selection in O4a.}
The joint profile-likelihood statistic $q_0$ is shown across the
$2$--$5~\kHz$ band for the clustered features.
Markers distinguish rejection by detector-response consistency,
the requirement of non-zero power in both detectors, and the noise
and morphology checks, together with the feature retained for
follow-up. The plotted statistic remains the joint $q_0$ at each stage.}
\label{fig:lpsd-q0-frequency}
\end{figure}

Figure~\ref{fig:lpsd-q0-frequency} summarizes these selections.
The additional follow-up of the surviving feature at $2597.723~\Hz$
is described in Section~\ref{sec:candidate-followup}.

\subsection{Candidate follow-up and validation}
\label{sec:candidate-followup}

No feature passes the final CC or BSD discovery selection. We nevertheless
investigate one sub-threshold feature from each method, together with the
single LPSD feature that survives the three candidate-selection cuts. The follow-up
tests examine persistence as data accumulate and consistency with the
finite spectral width of the adopted ULDM model. These are distinct
requirements: a persistent instrumental line can also strengthen with
observing time. The coherence tests must account for both the field
spectrum and the frequency response of the analysis window.

\subsubsection{Cross-correlation feature at $4931.8433~\Hz$}

The CC feature has $\mathrm{SNR}=-5.15+1.92i$ and passes the instrumental
and frequency-lag checks, but remains below the final discovery
threshold. We calculate its cumulative real SNR as synchronized SFT
pairs are accumulated and fit a dependence proportional to
$\sqrt{N_{\mathrm{SFT}}}$ (\fig{fig:snr-tobs}). The early data deviate
from the fitted trend; at larger accumulated durations, the statistic
fluctuates around it. This persistence check does not provide grounds
for rejecting the assumed signal interpretation.

We also compare the local cross-correlation spectrum with the expected
ULDM profile in \fig{fig:veto_summary}. The observed excess is narrower
than the reference signal profile, providing a separate spectral
consistency check. The formal non-detection already follows from the
discovery threshold; it does not depend on assigning a rejection
significance to this width comparison.

\begin{figure}[ht!]
\centering
\includegraphics[width=0.80\linewidth]{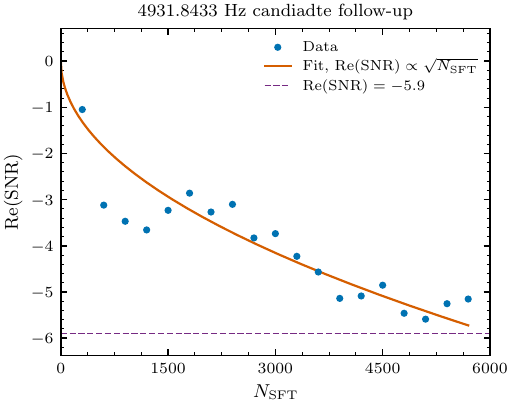}
\caption{\textbf{Cumulative cross-correlation SNR at
$4931.8433~\Hz$.} Blue circles show $\mathrm{Re}(\mathrm{SNR})$ as
synchronized SFT pairs are accumulated. The orange curve is a fit
proportional to $\sqrt{N_{\mathrm{SFT}}}$; the dashed line marks
$\mathrm{Re}(\mathrm{SNR})=-5.9$, the negative branch of the discovery
threshold. The cumulative values share data and are correlated. This
persistence diagnostic is considered together with the spectral-width
comparison.}
\label{fig:snr-tobs}
\end{figure}

\subsubsection{BSD feature at $2881~\Hz$}

The BSD feature has $CR=8.85$ in LHO and $CR=5.09$ in LLO. It passes
the frequency-coincidence and critical-ratio compatibility cuts, and
is retained for follow-up when the final threshold is relaxed to
$CR>5$. We first increase $\Tobs$ in approximately 10-day steps at
fixed $\TFFT$, recompute the critical ratio in each detector, and
compare the trends with fits proportional to $\sqrt{\Tobs}$.
The statistics strengthen with accumulated observing time, as shown
in the left panel of \fig{fig:cr-tfft-tobs}.

We then increase $\TFFT$ beyond its search value and compare the
critical ratios with the finite-coherence prediction used in the
BSD analysis~\cite{Miller:2020vsl,Miller:2022wxu}. Once a transform
resolves the field linewidth, signal power occupies several Fourier
bins, changing the response of a statistic based on individual selected
peaks. The observed feature continues to strengthen as $\TFFT$
increases, without the weakening predicted for the single-bin statistic
under the adopted ULDM model. This behaviour is consistent with a
persistent monochromatic instrumental feature.

\begin{figure}[ht!]
\centering
\subfigure[]{\label{fig:cr-all-bins}\includegraphics[width=0.48\linewidth]{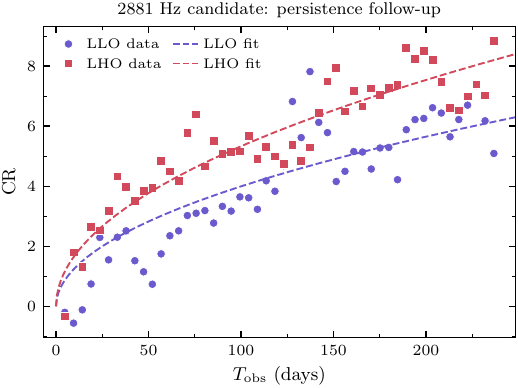}}
\hfill
\subfigure[]{\label{fig:cr-tail}\includegraphics[width=0.48\linewidth]{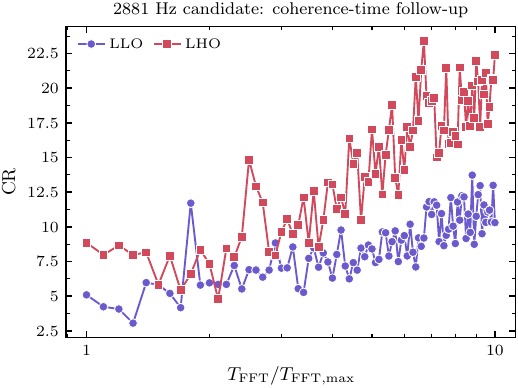}}
\caption{\textbf{Persistence and coherence-time checks for the BSD
feature at $2881~\Hz$.} Left, the critical ratio as the observing time
increases in steps of approximately 10 days at fixed $\TFFT$.
Right, the critical ratio as $\TFFT$ increases beyond its search value.
The continued strengthening with Fourier duration is compared with the
finite-coherence prediction for the adopted ULDM signal and is
consistent with a monochromatic instrumental feature.}
\label{fig:cr-tfft-tobs}
\end{figure}

\subsubsection{LPSD feature at $2597.723~\Hz$}

The LPSD feature has $q_0=60.8$ in LLO and $q_0=6.3$ in LHO. It
survives the local-noise, spectral-shape and relative-amplitude checks,
although the LHO statistic only slightly exceeds the local selection
threshold. We track the peak $\sqrt{q_0}$ within $\pm0.01~\Hz$ of the
candidate as data are accumulated separately for each detector and
compare the evolution with fits proportional to the square root of
the contributing strain time. The LLO evolution is compatible with
this dependence, whereas the LHO data are insufficient to establish it.
To limit computational cost, we compute each detector's LPSD spectrum
in a single configuration spanning 10~Hz--5~kHz, using a common data
selection determined by the Fourier duration
$T_{\mathrm{FFT}}\sim10^5~\mathrm{s}$ required near 10~Hz.
Data gaps limit the usable exposure at this long duration, so the LPSD
observing time is shorter than the total O4a time available after
data-quality cuts.
This loss is more severe in LHO, where more frequent gaps leave
substantially less usable data than in LLO
(\fig{fig:lpsd-q0-tobs}).

\begin{figure}[ht!]
\centering
\includegraphics[width=0.90\linewidth]{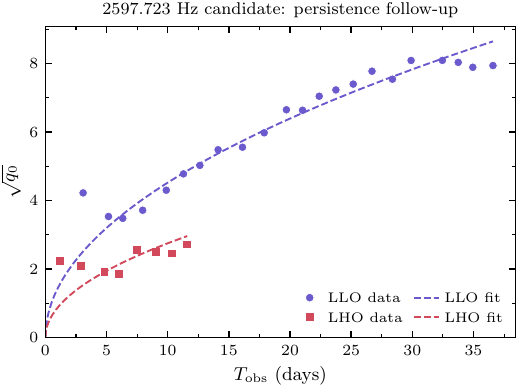}
\caption{\textbf{Cumulative LPSD statistic near $2597.723~\Hz$.}
Curves show the peak $\sqrt{q_0}$ within $\pm0.01~\Hz$ of the candidate
as a function of contributing strain time for the selections indicated
in the legend.
Dashed curves are fits
of the form $\sqrt{q_0}=a\sqrt{\Tobs}$, with $\Tobs$ in detector-days.
The quoted fit uncertainties are formal errors; cumulative subsets
share data and are correlated.}
\label{fig:lpsd-q0-tobs}
\end{figure}

We also vary the Fourier integration time to test the coherence of the
excess. The statistic remains elevated beyond the assumed field
coherence time, without the decrease predicted for the adopted
virialized ULDM signal. This is the coherence test used to reject the
remaining LPSD candidate under that signal model.

\subsubsection{Cross-pipeline comparisons and signal retention}

The three pipelines provide complementary tests over their respective
frequency ranges. Their coverage and the available statistics at the
three selected features are summarized in
\tab{tab:followup-crosschecks}. 

\begin{table}[ht]
\centering
\caption{Pipeline coverage and reported statistics at the selected
features. The entries distinguish evaluated statistics from coverage
and masking status. H and L denote LHO and LLO, respectively.}
\label{tab:followup-crosschecks}
\small
\setlength{\tabcolsep}{6pt}
\renewcommand{\arraystretch}{1.15}
\begin{tabularx}{\linewidth}{@{}l*{3}{>{\raggedright\arraybackslash}X}@{}}
\toprule
$f_0$ (Hz) & CC & BSD & LPSD \\
\midrule
4931.8433
& $\mathrm{SNR}=-5.15+1.92i$. Below the discovery threshold.
Excluded by line-width follow-up.
& Outside search range.
& $q_0\sim8.6$, but zero excess power identified in LHO, thus failing Cut 2. The morphology in LLO is broader than expected for \uldm and would also fail Cut 3. \\
2881
& Excluded by the instrumental-band candidate mask.
& $CR_{\mathrm H}=8.85$, $CR_{\mathrm L}=5.09$; sub-threshold and excluded by line-width follow-up.
& Sub-threshold $q_0\sim2.5$, and zero excess power identified in LLO. \\
2597.723
& SNR being far below the candidate threshold.
& Selected only in LLO with $CR_L=3.86$; not selected in H, thus failing Cut 1.
& $q_{0,\mathrm H}=6.3$, $q_{0,\mathrm L}=60.8$; survives Cut~3 but is excluded by line-width follow-up. \\
\bottomrule
\end{tabularx}
\end{table}

\clearpage

\backmatter

\section*{Acknowledgements}

This work is partly supported by the National Natural Science Foundation of China (NSFC) under Grant No. 12547104, 
W2611007,
and by the Fundamental Research Funds for the Central Universities.
The research of F.Y. is supported by the NSF Grant PHY-2412701 and the Cluster of Excellence $\textit{PRISMA}^{++}$ (EXC 2118/2, Project ID 390831469), funded by the German Research Foundation (DFG). Y.Z. acknowledges support from
the Research Grants Council of Hong Kong through the General Research Fund 16303026.
This material is based upon work supported by NSF's LIGO Laboratory, a major facility fully funded by the National Science Foundation.
The authors gratefully acknowledge the support of the NSF, STFC, INFN and CNRS for provision of computational resources.

\bibliography{sn-bibliography}

\end{document}